\documentclass[a4paper,oneside,final,notitlepage,onecolumn,12pt]{article}
\usepackage{amsfonts}
\usepackage{epsf}
\usepackage{graphicx}
\usepackage{amssymb}

\usepackage{t1enc}
\DeclareFontFamily{OT1}{rsfs}{} \DeclareFontShape{OT1}{rsfs}{m}{n}{
<-7> rsfs5 <7-10> rsfs7 <10-> rsfs10}{}
\DeclareMathAlphabet{\mycal}{OT1}{rsfs}{m}{n}

\def\sc{{\hskip 3.5pt {{}^{{}^{{}_{{}_{\bowtie}}}}} \kern -8.pt{}}}  
\def\SC{{\hskip 3.5pt {{}^{{}^{{}^{{}_{{}_{\bowtie}}}}}} \kern -10.5pt{}}}

\DeclareFontFamily{OT1}{rsfs}{} \DeclareFontShape{OT1}{rsfs}{m}{n}{
<-7> rsfs5 <7-10> rsfs7 <10-> rsfs10}{}
\DeclareMathAlphabet{\mathscr}{OT1}{rsfs}{m}{n}

\def\eeepsilon{{\epsilon}{\hskip-.1799cm\epsilon}}

\begin{document}

\newtheorem{theorem}{Theorem}[section]
\newtheorem{lemma}{Lemma}[section]
\newtheorem{proposition}{Proposition}[section]
\newtheorem{corollary}{Corollary}[section]
\newtheorem{conjecture}{Conjecture}[section]
\newtheorem{example}{Example}[section]
\newtheorem{definition}{Definition}[section]
\newtheorem{remark}{Remark}[section]
\newtheorem{exercise}{Exercise}[section]
\newtheorem{axiom}{Axiom}[section]
\renewcommand{\theequation}{\thesection.\arabic{equation}} 

\author{ Istv\'an R\'{a}cz\thanks{ 
 email: iracz@sunserv.kfki.hu, 
\newline 
This research was supported in part by OTKA grant K67942.}  
\\ 
RMKI, \\ H-1121
Budapest, Konkoly Thege Mikl\'os \'ut 29-33. \\Hungary
}
\title{A simple proof of the recent generalisations 
of Hawking's black hole topology theorem}   
\maketitle

\begin{abstract}   
A key result in four dimensional black hole physics, since the early 1970s, is
Hawking's topology theorem asserting that the cross-sections of an ``apparent
horizon'', separating the black hole region from the rest of the spacetime,
are topologically two-spheres. Later, during the 1990s, by applying a variant
of Hawking's argument, Gibbons and Woolgar could also show the existence of a
genus dependent lower bound for the entropy of topological black holes with
negative cosmological constant. Recently Hawking's black hole topology
theorem, along with the results of  Gibbons and Woolgar, has been generalised
to the case of black holes in higher dimensions. Our aim here is to give a
simple self-contained proof of these generalisations which also makes their
range of  applicability transparent.
\end{abstract}

\newpage

\section{Introduction}
\setcounter{equation}{0}
 
The notion of a trapped surface was introduced by Penrose \cite{p1}. In
$4$-dimensional spacetime the spacelike boundary, $\mathscr{S}$, of a
$3$-dimensional spatial region is called a future trapped surface if gravity
is so strong there that even the future and `outwards' directed normal null
rays starting at $\mathscr{S}$ are dragged back so much that their expansion
is non-positive everywhere on $\mathscr{S}$. Careful analysis justified that
trapped surfaces necessarily occur whenever sufficient amount of energy is
concentrated in a small spacetime region \cite{sy1}.    

Intuitively a black hole region is considered to be a part of a spacetime from
which nothing can escape. Therefore a black hole region is supposed to be a
future set comprised by events that individually belong to some future trapped
surface. The boundary of such a black hole region, referred to usually as the
``apparent horizon'', $\mathcal{H}$, is then supposed to be comprised by
marginal future trapped surfaces.  As one of the most important recent results
in black hole physics the existence of an ``apparent horizon'' was proved in
\cite{AMS,AMSn}. More specifically, it was shown there that given a strictly
stable {\it marginally outer trapped surface}  (MOTS)
$\mathscr{S}_0\subset\Sigma_0$ in a spacetime with reference foliation
$\{\Sigma_t\}$, then, there exists an open {\it tube},
$\mathcal{H}=\cup_t\mathscr{S}_t$, foliated by marginally outer trapped
surfaces, with $\mathscr{S}_t\subset\Sigma_t$, through $\mathscr{S}_0$. Let us
merely mention, without getting into details here, that the applied strict
stability assumption is to exclude the appearance of future trapped surfaces
in the complementer of a black hole region.

Hawking's black hole topology theorem \cite{hawk} is proven by demonstrating
that whenever  the dominant energy condition ({\it DEC}) holds a MOTS
$\mathscr{S}$ can be deformed, along  the  family of null geodesics transverse
to the apparent horizon, yielding thereby---on contrary to the fact that
$\mathscr{S}$ is a MOTS---a future trapped surface in the complementer of the
black hole region, unless the Euler characteristic $\chi_{_{\mathscr{S}}}$ of
$\mathscr{S}$ is positive. Whenever $\mathscr{S}$ is a codimension two surface
in a $4$-dimensional spacetime the Euler characteristic and the ``genus'',
$g_{_\mathscr{S}}$, of $\mathscr{S}$ can be given, in virtue of the
Gauss-Bonnet theorem,  via the integral of the scalar curvature $R_q$ of the
metric  $q_{ab}$ induced on $\mathscr{S}$ as  
\begin{equation}\label{tipp2e} 
2\pi\chi_{_\mathscr{S}}=4\pi(1-g_{_\mathscr{S}})=\frac12\int_\mathscr{S}R_q\,
\eeepsilon_q\,. 
\end{equation}

The main difficulty in generalising Hawking's argument to the higher
dimensional case originates from the fact that whenever $\mathscr{S}$ is of
dimension $s=n-2\geq 3$ in an $n$-dimensional spacetime, the integral of the
scalar curvature $R^{^{(s)}}$ by itself is not informative, as opposed to the
case of $n=4$, therefore the notion of Euler characteristic has to be replaced
by the Yamabe invariant. The latter is known to be a fundamental topological
invariant and it is defined as follows. Denote by  $[q]$ the conformal class of
Riemannian metrics on $\mathscr{S}$ determined by $q_{ab}$. It was conjectured
by Yamabe, and later proved by Trudinger, Aubin, and Schoen that to every
conformal class on any smooth compact manifold there exists a metric $\tilde
q_{ab}$ of constant scalar curvature so that 
\begin{equation}
R_{\tilde q}=Y(\mathscr{S},[q])\cdot \left(\int_{\mathscr{S}}\eeepsilon_{\tilde
  q}\right)^{-\frac{2}{s-2}}\,, 
\end{equation}
where the Yamabe constant $Y(\mathscr{S},[q])$, associated with the conformal
class $[q]$, is defined as 
\begin{equation}\label{y1}
Y(\mathscr{S},[q])= \inf_{\hat q\in[q]}\frac{\int_{\mathscr{S}}R_{\hat
    q}\,\eeepsilon_{\hat q}} 
{\left(\int_{\mathscr{S}}\eeepsilon_{\hat q}\right)^{\,\,\frac{s-2}{s}}}
=\inf_{u\in C^{\infty}(\mathscr{S}),u>0}\frac{\int_{\mathscr{S}}\left[ 
4\frac{s-1}{s-2}(D^au)(D_au)+R_qu^2
\right]\,\eeepsilon_{q}} 
{\left(\int_{\mathscr{S}}u^{\frac{2s}{s-2}}\eeepsilon_{q}\right)^{\,\,\frac{s-2}{s}}}
\,.
\end{equation} 
In the later case, the metric $\hat q\in[q]$ can be
given as $\hat q_{ab}= u^{\frac{4}{s-2}}\,q_{ab}$, and, moreover, $D_a$ and $R_q$
denote the covariant  derivative operator and the scalar curvature associated
with the metric $q_{ab}$ on $\mathscr{S}$.
The Yamabe invariant is defined then as 
\begin{equation}\label{y2}
\mathcal{Y}(\mathscr{S}) = \sup_{[q]}Y(\mathscr{S},[q]). 
\end{equation}

Some of the recent generalisation of Hawking's \cite{hawk} black hole topology
theorem, and also that of Gibbons' \cite{Gib} and Woolgar's \cite{wo} results,
proved by Galloway, Schoen, O'Murchadha and Cai, that are covered by
Refs.\,\cite{cg,ga1,ga3} may be formulated then as. 
\begin{theorem}\label{gs}
Let $(M,g_{ab})$ be a spacetime of dimension $n\geq 4$ satisfying the Einstein
equations
\begin{equation}\label{ein}
R_{ab}-\frac{1}{2} g_{ab} R+\Lambda g_{ab}=8\pi T_{ab},
\end{equation}   
with cosmological constant $\Lambda$, and with matter subject to {\it DEC}.
Suppose, furthermore, that $\mathscr{S}$ is a strictly stable 
MOTS in a regular spacelike hypersurface $\Sigma$.   
\begin{itemize}
\item[(1)] If $\Lambda\geq 0$ then $\mathscr{S}$ is of positive Yamabe type,
  i.e., $\mathcal{Y}(\mathscr{S})>0$. 
\item[(2)] If $\mathcal{Y}(\mathscr{S})<0$ and $\Lambda<0$ then for the
  ``area'' $\mathcal{A}(\mathscr{S})=\int_{\mathscr{S}}\,\eeepsilon_q$ of
  $\mathscr{S}$ the inequality 
\begin{equation}\label{gib}
\mathcal{A}(\mathscr{S})\geq
\left(\frac{|\mathcal{Y}(\mathscr{S})|}{2|\Lambda|}\right)^{\,\,\frac s2} 
\end{equation}
holds.
\end{itemize}     
\end{theorem} 

The significance of these results get to be transparent if one
recalls that in the first case, i.e., when $\mathcal{Y}(\mathscr{S})>0$,
$\mathscr{S}$ cannot carry a metric of non-positive sectional curvature which
immediately restricts the topological properties of $\mathscr{S}$
\cite{GL}. Whereas, in the second case the a lower bound on the ``entropy'' of
a black hole, that is considered to be proportional to the area
$\mathcal{A}(\mathscr{S})$, is provided by (\ref{gib}).

\bigskip

Before proceeding we would like to stress on an important conceptual
point. Most of the quoted investigations of black holes, see
Refs.\,\cite{AMS,AMSn,cg,ga1,ga2,ga3}, starts by assuming the existence of a
reference foliation $\{\Sigma_t\}$ of the spacetime by (partial) Cauchy
surfaces $\Sigma_t$. In this respect it is worth recalling that by a
non-optimal choice of $\{\Sigma_t\}$ one might completely miss a black hole
region as it follows from the results of \cite{wi} where it was demonstrated
that even in the extended Schwarzschild spacetime one may find a sequence of
Cauchy surfaces which get arbitrarily close to the singularity such that
neither of these Cauchy surfaces contains a future trapped surface. Hence, one
of the motivations for the present work---besides providing a reduction of the
complexity of the proof of Theorem\,\ref{gs}, and also a simultaneous widening
of its range of applicability---was to carry out a discussion without making
use of any reference foliation.

\section{Preliminaries}\label{stac}
\setcounter{equation}{0}

As it will be seen below the simplicity of the presented argument allows the
investigation of black holes essentially in arbitrary metric theory of
gravity. Thereby, we do not restrict our considerations to either of the
specific theories. Accordingly, a spacetime is assumed to be represented by a
pair $(M,g_{ab})$, where $M$ is an $n$-dimensional ($n\geq4$), smooth,
paracompact, connected, orientable manifold endowed with a smooth Lorentzian
metric $g_{ab}$ of signature $(-,+,\dots,+)$. It is assumed that $(M,g_{ab})$
is time orientable and that a time orientation has been chosen.

The only restriction, concerning the geometry of the allowed spacetimes, is
the following generalised version of {\it DEC}. A spacetime $(M,g_{ab})$ is
said to satisfy the {\it generalised dominant energy condition} if there
exists some smooth real function $f$ on $M$ such that for all future directed
timelike vector $t^a$ the combination $-[{G^a}_bt^b+f\,t^a]$ is a future
directed timelike or null vector, where $G_{ab}$ denotes the Einstein  tensor
$R_{ab}-\frac{1}{2} g_{ab} R$.  It is straightforward to see that in
Einstein's theory of gravity, where $g_{ab}$ is subject to (\ref{ein}), the
generalised dominant energy condition, with the choice $f=\Lambda$, is
equivalent to requiring the energy-momentum tensor, $T_{ab}$, to satisfy {\it
DEC} \cite{wald}.

\medskip

To restrict our considerations to black hole spacetimes, we shall also assume
the existence of future trapped surfaces in $(M,g_{ab})$ which are defined as
follows. Let us consider a smooth orientable $(n-2)$-dimensional compact
manifold $\mathscr{S}$ with no boundary in $M$. Let $\ell^a$ and $k^a$ be
smooth future directed null vector fields on $\mathscr{S}$ scaled such that
$k^a\ell_a=-1$, and that are also normal to $\mathscr{S}$, i.e.,
$g_{ab}\ell^aX^b|_{\mathscr{S}}=0$ and $g_{ab}k^aX^b|_{\mathscr{S}}=0$ for any
vector field $X^a$ tangent to $\mathscr{S}$. Consider then the null
hypersurfaces generated by geodesics starting on $\mathscr{S}$ with tangent
$\ell^a$ and $k^a$. These null hypersurfaces are smooth in a neighbourhood of
$\mathscr{S}$, and, by making use of the associated synchronised affine
parametrisations of their null generators, the vector fields $\ell^a$ and
$k^a$ can be, respectively, extended to them. The level surfaces
of the corresponding synchronised affine parametrisations do also provide
foliations of these null hypersurfaces by $(n-2)$-dimensional compact
manifolds homologous to $\mathscr{S}$. Denote by\, ${\eeepsilon}_q$ the volume
element associated with the metric, $q_{ab}$, induced on these
$(n-2)$-dimensional surfaces. Then the null expansions $\theta^{(\ell)}$ and
$\theta^{(k)}$ with respect to $\ell^a$ and $k^a$ are defined by 
\begin{equation}\label{exp}
\pounds_\ell\,\eeepsilon_q=\theta^{(\ell)}\,{\eeepsilon}_q
\,, \ \ \ 
\pounds_k\,{\eeepsilon}_q= \theta^{(k)}\,{\eeepsilon}_q\,,   
\end{equation}  where $\pounds_\ell$ and $\pounds_k$ denotes the Lie
derivative with respect to the null vector fields $\ell^a$ and $k^a$. 

The $(n-2)$-dimensional surface $\mathscr{S}$ is called future {\it trapped
surface}  if both of the null expansions $\theta^{(\ell)}$ and $\theta^{(k)}$
are non-positive on $\mathscr{S}$. In  the limiting case, i.e., whenever
either of these null expansions (say $\theta^{(\ell)}$) vanishes on
$\mathscr{S}$ identically, $\mathscr{S}$ is called future {\it marginally
trapped surface}. 

In case of a generic $(n-2)$-dimensional surface the quasi-local concept of
outwards and  inwards directions is undetermined. Nevertheless, these concepts
get to be well-defined for non-minimal marginally trapped surfaces. It will be
said that $\ell^a$  and $k^a$ point {\it outwards}  and {\it inwards},
respectively, provided that $\theta^{(\ell)}=0$ and $\theta^{(k)}\leq 0$, and
that $\theta^{(k)}$ is not identically zero. (If both $\theta^{(\ell)}$ and
$\theta^{(k)}$ vanish identically $\mathscr{S}$ is a minimal surface and the
notions outwards and inwards become to be degenerate.)      

\begin{figure}[htbp!]
 \centerline{
  \epsfxsize=7cm 
\epsfbox{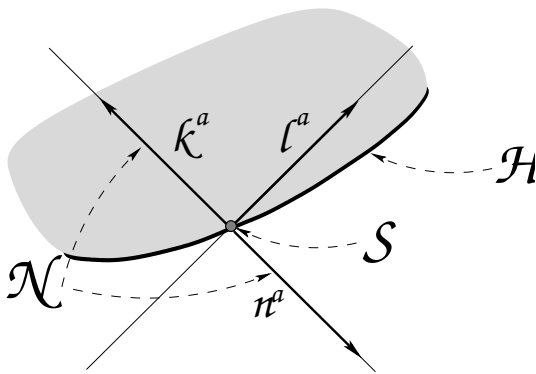}
 }
\caption{\footnotesize \label{mots} The black hole, represented by the shaded
  region, is bounded by horizon $\mathcal{H}$ that is foliated by MOTS'
  homologous to $\mathscr{S}$.}
\end{figure}

To see that this, quasi-local, concept of ``outer'' direction is not counter
intuitive, consider the null hypersurface $\mathcal{N}$ generated by null
geodesics starting at the points of $\mathscr{S}$ with tangent $n^a=-k^a$ (see
Fig.\,\ref{mots}). 
Since $\mathcal{N}$ is smooth in a neighbourhood $\mathcal{O}$ of $\mathscr{S}$
it can be smoothly foliated by $(n-2)$-dimensional surfaces, $\mathscr{S}_u$,
defined as the $u=const$ cross-sections of $\mathcal{N}$, where $u$ is the
affine parameter along the generators of $\mathcal{N}$ such that
$n^a=(\partial/\partial u)^a$ and $u=0$ on $\mathscr{S}$. Then, it seems to be
natural to consider $k^a$ as inward pointing if the ``area''
$\mathcal{A}(\mathscr{S}_u)=\int_{\mathscr{S}_u}\,\eeepsilon_q$ of
the cross-sections $\mathscr{S}_u$ is non-decreasing in the
direction of $n^a$ which, in the case under consideration, follows from
$\theta^{(k)}\leq 0$ as  
\begin{equation}\label{av00}
\frac{{\rm d} \mathcal{A}(\mathscr{S}_u)}{{\rm
    d}u}\vert_{u=0}= \int_\mathcal{\mathscr{S}}
    \pounds_{-k}\,\eeepsilon_q=
    -\int_\mathcal{\mathscr{S}} \theta^{(k)}\,\eeepsilon_q\geq 0\,.
\end{equation} 

Accordingly, $\mathscr{S}$ is called future {\it marginally outer trapped
surface} (MOTS) if $\theta^{(\ell)}=0$ and $\theta^{(k)}\leq 0$ on
$\mathscr{S}$.  

\smallskip

In deriving our results we shall also apply a stability assumption. Before
formulating it let us recall first that the above imposed conditions do not
uniquely determine the pair of null vector fields $\ell^a$ and $k^a$ on
$\mathscr{S}$. In fact, together with $\ell^a$ and $k^a$ any pair of null
vector fields $\ell'^a$ and $k'^a$ that is yielded by the boost transformation 
\begin{equation}\label{boost}
k^a\rightarrow k'^a=e^v\,k^a\,, \ \  \ell^a\rightarrow
\ell'^a=e^{-v}\,\ell^a\,, 
\end{equation}  where $v: \mathscr{S} \rightarrow \mathbb{R}$ is an arbitrary
smooth function on $\mathscr{S}$, will be suitable. It
is well-known, however, that the signs of $\theta^{(\ell)}$ and $\theta^{(k)}$
are invariant under such positive rescaling of $\ell^a$ and $k^a$. 

Suppose then that $\mathscr{S}$ is a future MOTS with respect to a null normal
$\ell^a$. Then, $\mathscr{S}$ will be called {\it  strictly stably outermost}
if there exists a rescaling of the type (\ref{boost}) so that $\pounds_{k'}\,
\theta^{(\ell')}\leq 0$ for the yielded vector fields $\ell'^a$ and $k'^a$,
and also $\pounds_{k'}\, \theta^{(\ell')}< 0$ somewhere on $\mathscr{S}$.
Obviously, this definition is independent of the use of any sort of reference
foliation. Moreover, as it will be indicated below, it can be seen to be
equivalent to the corresponding criteria applied in \cite{AMS,AMSn}.

\section{The proof of the main result}\label{proof}
\setcounter{equation}{0}

The main argument of the present paper can then be given in the following
simple geometric setup. We have already defined $n^a=-k^a$ to be a smooth past
directed null vector field on $\mathscr{S}$ that is also normal to
$\mathscr{S}$. Similarly, the smooth null hypersurface $\mathcal{N}$, spanned
by the $(n-2)$-parameter congruence of null geodesics starting at
$\mathscr{S}$ with tangent $n^a$, and the affine parameter $u$ along the
geodesics, synchronised such that $u=0$ on  $\mathscr{S}$, have already been
introduced. Denote by $n^a$ the tangent field $(\partial/\partial u)^a$ on the
null  hypersurface $\mathcal{N}$, that is foliated by the smooth $u=const$
cross-sections, $\mathscr{S}_u$. Then, there exists a uniquely defined future
directed null vector field $\ell^a$ on $\mathcal{N}$ such that
$g_{ab}n^a\ell^b=1$, and that $\ell^a$ is orthogonal to each
$\mathscr{S}_u$. Denote by $r$ the affine parameter of the null geodesics
determined by $\ell^a$ which are synchronised such that $r=0$ on
$\mathcal{N}$.  

\medskip

Since $\ell^a$ is, by construction, smooth on $\mathcal{N}$ the null geodesics
starting with tangent $\ell^a$ on $\mathcal{N}$ do not meet  within a
sufficiently small open  ``elementary spacetime neighbourhood'' $\mathcal{O}$
of $\mathscr{S}$. Extend, then, the function $u$ from $\mathcal{N}$ onto
$\mathcal{O}$ by keeping its value to be constant along the geodesics with
tangent  $\ell^a$. Then the vector fields $n^a$ and $\ell^a$, defined so far
only on $\mathcal{N}$, do also extend onto $\mathcal{O}$ such that the relations
$n^a=(\partial/\partial u)^a$ and $\ell^a=(\partial/\partial r)^a$ hold there
which do also imply  that $n^a$ and $\ell^a$ commute on $\mathcal{O}$. Note
that $\mathcal{O}$ is smoothly foliated by the $2$-parameter
family of $(n-2)$-dimensional  $u=const$, $r=const$ level surfaces
$\mathscr{S}_{u,r}$. The spacetime metric in $\mathcal{O}$ can then always be
given (see, e.g., \cite{hiw} for a justification) as 
\begin{equation}\label{metric}
g_{ab}=2\,\left(\nabla_{(a}r - r\,\alpha\,\nabla_{(a}u - r\,\beta_{(a}\right)
\nabla_{b)}u +\gamma_{ab} 
\end{equation} where $\alpha$, $\beta_a$ and $\gamma_{ab}$ are smooth 
fields on $\mathcal{O}$ such that $\beta_a$ and $\gamma_{ab}$ are orthogonal to
$n^a$ and $\ell^a$.  Recall also that $\gamma_{ab}$ and the positive definite
metric $q_{ab}$ on the $(n-2)$-dimensional surfaces $\mathscr{S}_{u,r}$ are
related as 
\begin{equation}\label{indmetric}
q_{ab}= r^2\,\beta^c\beta_c \,{\ell}_a\ell_b
-2\,r\beta_{(a}\ell_{b)}+\gamma_{ab}\,.  
\end{equation}
This latter relation implies that $q_{ab}=\gamma_{ab}$ on $\mathcal{N}$,
represented by the $r=0$ hypersurface in $\mathcal{O}$, 
i.e., $\gamma_{ab}$ is the metric on the cross-sections $\mathscr{S}_{u}$ of
$\mathcal{N}$. 

Since the vector fields $n^a$ and $\ell^a$ are null and normal to the
cross-sections $\mathscr{S}_{u}$ the expansions of the associated null
congruences at $\mathscr{S}_{u}$ can be given as 
\begin{equation}\label{exp2}
\theta^{(n)}|_{\mathscr{S}_{u}}=\frac12 q^{ef}\left(\pounds_n
q_{ef}\right)=\frac12 \gamma^{ef}\left(\pounds_n
\gamma_{ef}\right)\ \ \ {\rm and} \ \ \ 
\theta^{(\ell)}|_{\mathscr{S}_{u}}=\frac12 q^{ef}\left(\pounds_\ell
q_{ef}\right)=\frac12 \gamma^{ef}\left(\pounds_\ell
\gamma_{ef}\right)\,, 
\end{equation}
where $\pounds_n$ and $\pounds_\ell$ denote the Lie derivative with respect to
the vector fields $n^a$ and $\ell^a$, respectively, and (here and elsewhere)
all the indices are raised and lowered with the spacetime metric $g_{ab}$. The
second equalities in (\ref{exp}) follow from the fact that $\beta_a$ and
$\gamma_{ab}$ are orthogonal to $n^a$ and $\ell^a$ and also that $n^a$ and
$\ell^a$ commute in $\mathcal{O}$.    

\bigskip

By making use of (\ref{metric}), and the definition of the Einstein tensor, it
is straightforward to see that   
\begin{equation}
G_{ab}n^a\ell^b=R_{ab}n^a\ell^b-\frac12
R_{ef}\left[ 2\,n^{(e}\ell^{f)} +\gamma^{ef}\right]= -\frac12 \gamma^{ef}R_{ef} 
\end{equation} 
holds on $\mathcal{N}$. Then, in virtue of equation $(82)$ of \cite{hiw}, and
by the coincidence of  $q_{ab}$ and $\gamma_{ab}$, and also of the projectors
${q^a}_{b}$, ${\gamma^a}_{b}$ and ${p^a}_{b}$ on $\mathcal{N}$ (for the
definition of ${p^a}_{b}$ see $(76)$ of \cite{hiw}) we also have that on
$\mathcal{N}$
\begin{eqnarray}\label{GDEC1}
&&G_{ab}n^a\ell^b=-\frac12\left[
-\gamma^{ab}\left(\pounds_\ell\pounds_n\gamma_{ab}\right) -\alpha\,
\gamma^{ab}\left(\pounds_\ell\gamma_{ab}\right) + R_q +D^a\beta_a-\frac12
\beta^a\beta_a\right.\nonumber \\
&&\left.\phantom{G_{ab}n^a\ell^b=-\frac12\,\,\,}+
\gamma^{ab}\gamma^{cd}\left(\pounds_\ell\gamma_{ac}\right)
\left(\pounds_n\gamma_{bd}\right) -\frac12 \gamma^{ab}\left(\pounds_\ell
\gamma_{ab}\right) \gamma^{cd}\left(\pounds_n \gamma_{cd}\right) \right]\,,  
\end{eqnarray} 
where $D_a$ and $R_q$ denote the covariant derivative operator
and the  scalar curvature associated with the metric $q_{ab}=\gamma_{ab}$ on
the $(n-2)$-dimensional surfaces $\mathscr{S}_{u}$ on $\mathcal{N}$. By making
use of the fact that the vector fields $n^a$ and $\ell^a$  commute in
$\mathcal{O}$ and that they are orthogonal to $\gamma_{ab}$ a direct
calculation justifies then the relation
\begin{equation}\label{sGDEC1}
-\gamma^{ab}\left(\pounds_\ell\pounds_n\gamma_{ab}\right)=
-\pounds_\ell\left(\gamma^{ab}\pounds_n\gamma_{ab}\right) -
\gamma^{ab}\gamma^{cd}\left(\pounds_\ell\gamma_{ac}\right) 
\left(\pounds_n\gamma_{bd}\right)\,.
\end{equation}

Since $n^a$ and $\ell^a$ commute we also have that
\begin{equation}\label{n}
\pounds_\ell\theta^{(n)}=\pounds_n\theta^{(\ell)}
\end{equation} 
in $\mathcal{O}$, where
the expansion $\theta^{(n)}$ is defined with respect to the volume element
$\eeepsilon_q$ associated with the metric $q_{ab}$ on the surfaces
$\mathscr{S}_{u,r}$ as in (\ref{exp}).  Similarly, it can be verified that
\begin{equation}\label{tk}
\pounds_k\theta^{(\ell)}=\pounds_n\theta^{(\ell)}
\end{equation}
on $\mathcal{N}$, i.e., whenever $r=0$, where
$k^a$ denotes the unique future directed null extension
\begin{equation}\label{k}
k^a=-\left[n^a+\left(r\alpha+\frac12 r^2\beta^e\beta_e\right) \ell^a + r
  \beta^a\right] 
\end{equation}
of $k^a=-n^a$ on $\mathscr{S}$ onto $\mathcal{O}$ which is normal to
the surfaces $\mathscr{S}_{u,r}$ and is scaled such that $k^a\ell_a=-1$ in
$\mathcal{O}$. 

Then, by making use of the vanishing of $\theta^{(\ell)}$ on $\mathscr{S}$,
the above relations---in particular, equations (\ref{exp}), (\ref{GDEC1}),
(\ref{sGDEC1}) and (\ref{n})---imply that
\begin{equation}\label{GDEC2}
\pounds_n\theta^{(\ell)}\vert_{\mathscr{S}}=
G_{ab}n^a\ell^b+\frac12\left[R_q+D^a\beta_a-\frac12\beta^a\beta_a
  \right]\,.  
\end{equation}  Since $-n^a$ and $\ell^a$ are both future directed null vector
fields on $\mathcal{N}$, and also the generalised dominant energy condition
holds, i.e., there exists a real function $f$ on $M$ such that the vector
field $-[{G^a}_bl^b+f\,l^a]$ is future directed and causal, the inequality
$G_{ab}n^a\ell^b+f\leq 0$ holds on $\mathcal{N}$. Finally, since $\mathscr{S}$
was assumed to be a strictly stable MOTS, in virtue of (\ref{tk}), the null
normals $n^a=-k^a$ and $\ell^a$ may be assumed, without loss of generality, to
be such that $\pounds_n\theta^{(\ell)}\vert_{\mathscr{S}}\geq 0$, and also
that $\pounds_n\theta^{(\ell)}\vert_{\mathscr{S}}> 0$ somewhere on
$\mathscr{S}$. 

To see that the stability condition applied here is equivalent to the one
used in \cite{AMS,AMSn} note that $\beta_a=-{q^e}_an_b\nabla_e\ell^b$ and it
transforms under the rescaling (\ref{boost}) of the vector fields $k^a=-n^a$
and $\ell^a$ on $\mathscr{S}$ as $\beta_a\rightarrow \beta'_a= \beta_a + D_a
v$. By making use of the notation $\psi=e^{-2v}$ and $s_a=\frac12\beta_a$, it
can be verified then that 
\begin{equation}\label{GDEC2a}
\pounds_{\psi n'}\theta^{(\ell')}\vert_{\mathscr{S}}=
-D^aD_a\psi+2s^aD_a\psi+\frac\psi2 \left[R_q+2 G_{ab}n^a\ell^b+2
  D^as_a-2s^as_a \right]\,  
\end{equation} 
holds, which is exactly the expression `$\delta_q\theta$' given in
Lemma\,3.1 of \cite{AMSn} whenever $\mathscr{S}$ is a MOTS and the variation
vector field $q^a$ is chosen to be `$\psi n^a$'. This justifies then that
the strict stability conditions applied here and in \cite{AMS,AMSn} (see,
e.g., Definition 5.1 and the discussion at the end of Section 5 of
\cite{AMSn} for more details) are  equivalent.    

\medskip

In returning to the main stream of our argument note that whenever
$\mathscr{S}$ is a strictly stable MOTS and the generalised {\it DEC} holds
then, in virtue of (\ref{GDEC2}),   
\begin{equation}\label{GDEC3}
R_q+ D^a\beta_a - \frac12 \beta^a\beta_a \geq 2\,f\,,
\end{equation}
so that the inequality is strict somewhere on $\mathscr{S}$. Since $q_{ab}$ is
positive definite we also have that for any smooth function $u$ on
$\mathscr{S}$ 
\begin{equation}\label{GDEC4}
u^2\,D^a\beta_a=D^a(u^2\beta_a)-2\,u(D^au)\beta_a\leq 
D^a(u^2\beta_a)+2 (D^au)(D_au)+\frac12 u^2\beta^a\beta_a\,.
\end{equation}
Thus, multiplying (\ref{GDEC3}) by $u^2$, where $u>0$ is arbitrary, we get, in
virtue of (\ref{GDEC4}), that 
\begin{equation}\label{GDEC5}
2 (D^au)(D_au) + R_q u^2+ D^a(u^2\beta_a) \geq 2\,fu^2\,,
\end{equation} 
so that the inequality is strict somewhere on $\mathscr{S}$.

\bigskip     

To get the analogue of the first part of Theorem\,\ref{gs} assume now that $f$
is such that $f\geq 0$ throughout $\mathscr{S}$. Then, by taking into account
the inequality 
$4\frac{s-1}{s-2}>2$, which holds for any value of $s\geq 3$, we get from
(\ref{GDEC5}) that  
\begin{equation}
\frac{\int_{\mathscr{S}}\left[ 4\frac{s-1}{s-2}(D^au)(D_au)+R_qu^2
\right]\,\eeepsilon_{q}} 
{\left(\int_{\mathscr{S}}u^{\frac{2s}{s-2}}\eeepsilon_{q}\right)^{\,\,\frac{s-2}{s}}}
>0\,, 
\end{equation} 
for any smooth $u>0$, i.e., $Y(\mathscr{S},[q])>0$, which implies that
$\mathscr{S}$ is of positive Yamabe type. 

Similarly, whenever the minimal value $f^{_{\mathscr{S}}}_{min}$ of $f$ on 
${\mathscr{S}}$ is negative, on one hand,  
\begin{equation}\label{c1}
2\int_{\mathscr{S}} f u^2\,\eeepsilon_{q}\geq -2
|f^{_{\mathscr{S}}}_{min}|\int_{\mathscr{S}}u^2\,\eeepsilon_{q}  
\end{equation}
while, on the other hand, by applying the H\"older inequality 
\begin{equation}
\int_{\mathscr{S}}\phi_1\phi_2\,\eeepsilon_{q}\leq
\left(\int_{\mathscr{S}}|\phi_1|^a\eeepsilon_{q}\right)^{\frac1a}
\left(\int_{\mathscr{S}}|\phi_2|^b\eeepsilon_{q}\right)^{\frac1b}\,, \ \ \
{\frac1a}+{\frac1b}=1 
\end{equation}
to the functions $\phi_1=u^2$ and $\phi_2=1$ with $a=\frac{s}{s-2}$, $b=\frac
s2$, we get that  
\begin{equation}\label{c2}
\int_{\mathscr{S}}u^2\,\eeepsilon_{q}\leq
\left(\int_{\mathscr{S}}u^{\frac{2s}{s-2}}\eeepsilon_{q}\right)^{\,\,\frac{s-2}{s}}
\left[ {\mathcal{A}}({\mathscr{S}}) \right]^{\,\,\frac{2}{s}}\,.  
\end{equation}
The combination of (\ref{c1}) and (\ref{c2}), along with (\ref{GDEC5}),
justifies then that  
\begin{equation}\label{c3}
\frac{\int_{\mathscr{S}}\left[ 
4\frac{s-1}{s-2}(D^au)(D_au)+R_qu^2
\right]\,\eeepsilon_{q}} 
{\left(\int_{\mathscr{S}}u^{\frac{2s}{s-2}}\eeepsilon_{q}\right)^{\,\,\frac{s-2}{s}}} 
\geq -2|f^{_{\mathscr{S}}}_{min}|\left[ {\mathcal{A}}({\mathscr{S}})
\right]^{\,\,\frac{2}{s}}  
\,.
\end{equation} 
Assuming finally that ${\mathcal{Y}}(\mathscr{S})<0$ we get that, for any
conformal class [q], the Yamabe constant $Y(\mathscr{S},[q])\leq
{\mathcal{Y}}(\mathscr{S})<0$. This, along with (\ref{c3}), implies then 
\begin{equation}
|{\mathcal{Y}}(\mathscr{S})|\leq |Y(\mathscr{S},[q])|= -Y(\mathscr{S},[q])
\leq  2 |f^{_{\mathscr{S}}}_{min}| \left[ {\mathcal{A}}({\mathscr{S}})
\right]^{\,\,\frac{2}{s}}\,,  
\end{equation}
which leads to the variant of the inequality (\ref{gib}) yielded by the
replacement of $\Lambda$ by $f^{_{\mathscr{S}}}_{min}$.

\section{Final remarks}\label{fr}
\setcounter{equation}{0}

What has been proven in the previous section can be summarised as.

\begin{theorem}\label{gsn}  Let $(M,g_{ab})$ be a spacetime of dimension
$n\geq 4$ in a metric theory of gravity. Assume that the generalised dominant
energy condition, with smooth real function $f:M\rightarrow \mathbb{R}$, holds
and that $\mathscr{S}$ is a strictly stable MOTS in $(M,g_{ab})$.
\begin{itemize}
\item[(1)] If $f\geq 0$ on $\mathscr{S}$ then $\mathscr{S}$ is of positive
  Yamabe type, 
  i.e., $\mathcal{Y}(\mathscr{S})>0$. 
\item[(2)] If $\mathcal{Y}(\mathscr{S})<0$ and $f^{_{\mathscr{S}}}_{min}<0$,
  where $f^{_{\mathscr{S}}}_{min}$ denotes the minimal value of $f$ on
  $\mathscr{S}$,  then  
\begin{equation}\label{gib2}
\mathcal{A}(\mathscr{S})\geq
\left(\frac{|\mathcal{Y}(\mathscr{S})|}{2|f^{_{\mathscr{S}}}_{min}|}\right)^{\,\,\frac
  s2}\,.  
\end{equation}
\end{itemize}     
\end{theorem} 

We would like to mention that the argument of the previous section does also
provide an immediate reduction of the  complexity of the original proof of
Hawking and that of Gibbons and Woolgar. To see this recall that, in virtue of
(\ref{tipp2e}), (\ref{y1}) and (\ref{y2}),  
$\mathcal{Y}(\mathscr{S})=4\pi \chi_{_{\mathscr{S}}}\,,$ 
whenever $\mathscr{S}$ is of dimension $s=2$, and also that
$f=\Lambda$, if attention is restricted to Einstein's theory with matter
satisfying the dominant energy condition. Thereby, as an immediate consequence
of Theorem\,\ref{gsn}, we have that 
\begin{equation}
\chi_{_{\mathscr{S}}}>0\,, 
\end{equation}
whenever $\Lambda\geq 0$ on $\mathscr{S}$, while
\begin{equation}\label{gib20}
\mathcal{A}(\mathscr{S})\geq
\frac{4\pi(1-g_{_{\mathscr{S}}})}{|\Lambda|} 
\end{equation}
whenever both $\chi_{_{\mathscr{S}}}$ and $\Lambda$ are negative.  

\smallskip

Clearly the above justification of Theorem\,\ref{gsn} is free of the use of
any particular reference foliation of the spacetime. 
Note also that in the topological characterisation of an $(n-2)$-dimensional
strictly stable MOTS  $\mathscr{S}$ only the
quasi-local properties of the real function $f:M\rightarrow \mathbb{R}$ are
important. 
In particular, as the conditions of Theorem\,\ref{gsn} do merely refer
to the behaviour of $f$ on $\mathscr{S}$ it need not to be bounded or to have
a characteristic sign throughout $M$. Similarly, it would suffice to require
the generalised dominant energy condition to be satisfied only on
$\mathscr{S}$.  

\smallskip

Finally, we would also like to emphasise that Theorem\,\ref{gsn} provides a
considerable widening of the range of applicability of the
generalisation of Hawking's black hole topology theorem, and also that of the
results of Gibbons and Woolgar. As its conditions indicate, Theorem\,\ref{gsn}
applies to any metric theory of gravity and the only restriction concerning
the spacetime metric is manifested by the generalised dominant energy
condition and by the assumption requiring the existence of a strictly stable
MOTS. Accordingly, Theorem\,\ref{gsn} may immediately be applied in string
theory or in various other higher dimensional generalisations of
general relativity.


\end{document}